\documentclass[10pt,onecolumn,a4paper]{article}
\newcommand{\affil}[1]{$^{\rm #1}$}
\usepackage[authoryear]{natbib}
\bibpunct{(}{)}{;}{a}{}{,}
\usepackage{graphicx}
\usepackage{psfig}
\usepackage{subfigure}
\date{} %Please leave the date blank
\title{Evaluation of Cosmic Ray Rejection Algorithms on Single-shot Exposures}
\author{{\it Catherine L.\ Farage\affil{A,B} and Kevin A.\ Pimbblet\affil{A,C}}\\\\
\affil{A}\,Department of Physics, University of Queensland, Brisbane, 4072 Queensland, Australia\\
\affil{B}\,email: s3362906@student.uq.edu.au\\
\affil{C}\,email: pimbblet@physics.uq.edu.au}
\begin{document}
 \maketitle
\begin{minipage}{.9\textwidth}
{\bf Abstract}\\
To maximise data output from single-shot astronomical images, the rejection of cosmic rays is 
important. We present the results of a benchmark trial comparing various cosmic ray rejection 
algorithms. The procedures assess relative performances and characteristics of the processes in 
cosmic ray detection, rates of false detections of true objects and the quality of image cleaning 
and reconstruction. The cosmic ray rejection algorithms developed by Rhoads (2000), van Dokkum 
(2001), Pych (2004) and the IRAF task \textsc{xzap} by Dickinson are tested using both simulated 
and real data. 
It is found that detection efficiency is independent of the density of cosmic rays in an image, 
being more strongly affected by the density of real objects in the field. As expected, spurious 
detections and alterations to real data in the cleaning process are also significantly increased 
by high object densities. 
We find the Rhoads' linear filtering method to produce the best performance in detection of cosmic ray 
events, however, the popular van Dokkum algorithm exhibits the highest overall performance in terms 
of detection and cleaning.

\medskip
\end{minipage}

%
% Body of paper
%
\section{Introduction}

Telescope observations suffer from the isotropic influx of (hadronic) 
cosmic rays striking their charge coupled device (CCD) detectors.  In imaging these 
cosmic rays manifest as features with highly significant power at spatial frequencies 
($\sim$one--several pixels) too high to be considered as legitimate objects (Rhoads 2000).  
Although there exist some uses for such cosmic rays (e.g.\ Pimbblet \& Bulmer (2005) use 
cosmic rays in the generation of random numbers), generally one will want to detect and 
remove them.

Cosmic rays accumulate linearly in time on a detector and significant numbers are 
accumulated on an image or spectrum during longer exposures. Particularly when 
obtained in the higher radiation environment in orbit or outside the Earth's 
magnetic field, images may suffer extensive data loss. For example, Offenberg et 
al. (1999) predict that up to 10 per cent of the {\it Next Generation Space Telescope}
field of view may be affected by cosmic rays during a 1000s exposure. Efficient 
mechanisms for detecting and removing such defects are a vital consideration.

There are a plethora of approaches to the problem of cosmic ray rejection (CRR 
herein).  The most frequently used solution is to have multiple images of the same 
astronomical target: cosmic rays are very unlikely to hit the same pixel more than once 
in a series of exposures.  The images in which the cosmic rays are not present 
are then used to compute a replacement value (Shaw \& Horne 1992; Windhorst, 
Franklin \& Neuschaefer 1994; Zhang 1995; Freudling 1995; Fruchter \& Hook 1997).

A tougher CRR problem arises in the case of single-shot exposures (e.g.\ for fast 
moving targets; time-critical observations or data processing; or where multiple 
images are simply unavailable; Pimbblet \& Drinkwater 2004). 
In these instances image processing techniques must exploit the 
intrinsic properties of CCD cosmic ray events: their sharpness and high power at 
high spatial frequency. 
A number of methods have been developed to identify and replace pixels affected by cosmic rays 
in both imaging and spectroscopic data. Some of these include median or mean 
filtering (e.g.\ Dickinson's IRAF tasks \textsc{qzap}, \textsc{xzap} and 
\textsc{xnzap}; http://www.starlink.ac.uk/iraf/ftp/iraf/extern-v212/xdimsum/xdimsum.readme), 
applying a threshold on the contrast of high intensity objects 
with surrounding pixels (e.g.\ IRAF task \textsc{cosmicrays}), convolution with 
specially designed adaptations of a point-spread function (PSF; Rhoads 2000), 
Laplacian edge detection (van Dokkum 2001), trainable classification of objects 
in an image (Murtagh 1992; Salzberg et al. 1995; {\sc SExtractor} by Bertin \& Arnouts, 
1996) and analysis of the image data histogram (Pych 2004). 

In implementing such a process, a maximum possible number of cosmic rays should be detected, 
flagged and `cleaned' (replaced by a reasonable interpolated pixel value). Coincidently, 
the method must minimise the number of non-cosmic ray pixels that are falsely flagged to 
prevent loss of useful data. 

This work presents an investigation into the performance of several CRR techniques 
for single-shot astronomical image processing. The primary aims are to make a comparison 
of the selected techniques and to examine their performance under different conditions. 
In Section~\ref{sec:eval} we present the rejection algorithms, details of the real and 
artificial datasets to which they are applied and the measures of performance used. 
The results of the tests are presented in Section~\ref{sec:results}, followed 
by a discussion of the results and the subsequent conclusions of the study in 
Section~\ref{sec:conc}.

\section[]{Evaluation}
\label{sec:eval}
\subsection{Algorithms}
Of the available cosmic ray detection and rejection algorithms, the following four commonly 
used methods were selected for testing (identified by the developer and algorithm name): 
Rhoads' (2000) IRAF script, \textsc{jcrrej2}; van Dokkum's (2001) IRAF script, 
\textsc{L.A.Cosmic}; Pych's (2004) C script, \textsc{dcr}; and Dickinson's IRAF task, 
\textsc{xzap} (http://www.starlink.ac.uk/iraf/ftp/iraf/extern-v212/xdimsum/xdimsum.readme).  
All of these algorithms are idempotent: repeat usage of them produces no further effect.  
In implementing each algorithm, the user provides the input to a number of free parameters 
that customise the behaviour of the process. These must, by necessity, be adjusted 
experimentally to obtain the best detection and reconstruction for a given image.
Here we briefly review the pertinent details of the selected algorithms and comment on 
the omission of several potential candidates. 

Rhoads' (2000) IRAF task implements a linear filtering process in which the free 
parameters are the image PSF and noise properties, the upper sigma clipping threshold 
for cosmic ray identification and a choice of the algorithm used to replace flagged 
pixels. The process convolves the image with a function derived from the 
difference between a Gaussian PSF and a scaled delta function. Pixels constituting 
cosmic ray events are then identified as those with intensities lying above 
the threshold value. The algorithm performs multiple iterations of the process to 
ensure that pixels `shielded' by cosmic ray neighbours in previous applications 
can detected. Rhoads (2000) indicates that the probability of the algorithm 
falsely flagging a pixel containing legitimate object flux as a cosmic ray will 
not exceed that for a blank sky pixel, in order to decrease the rejection of true data. 
This safety check is likely to cost some efficiency in rejecting cosmic rays on real objects.  
The point is also made that well sampled data with, in practice, a seeing of $\ge$
2 pixels is required for successful operation of the algorithm (Rhoads 2000).
The output of the algorithm is a cosmic ray mask image in which only the cosmic ray affected 
pixels are assigned non-zero flux values. The Rhoads' method does not directly output 
a cleaned image from which the detected cosmic rays have been removed and are replaced. An 
independent algorithm is required in order to complete the cosmic ray rejection process 
(for example, the IRAF \textsc{fixpix} task), though this has not been performed in these tests
of the capabilities of the cosmic ray rejection algorithms.

The L.A.Cosmic -- Laplacian Cosmic Ray Identification -- algorithm (van Dokkum 2001) 
relies on the sharpness of the edges of cosmic ray objects in the detection process. 
The free parameters are: a detection limit for cosmic rays in terms of the 
background standard deviation ($\sigma$), a criterion for the detection of 
neighbouring pixels and a contrast limit for discriminating between cosmic 
rays and other objects. Again, the process is iterative, and at each step makes 
an estimate of the image noise properties and gain (if not provided at the input), 
identifies cosmic ray pixels, differentiates between cosmic rays and objects and 
fixes the flagged pixels. When no further cosmic ray candidates are detected, the 
iterations halt and the cleaned image and a bad pixel map of the detected objects 
are output. The algorithm has been designed with the intention of application to 
potentially undersampled data from the Hubble Space Telescope (HST), and the author
makes recommendations as to appropriate example settings for such images.

Pych (2004) takes a different approach to cosmic ray detection. Rather than using 
an image filtering or modeling technique, the algorithm makes an analysis of the 
histogram of counts in small sub-sections of the image. The free parameters include 
a threshold value, the size of the sub-frame box and details about the method of 
interpolation and replacement of flagged pixels. The user can also specify a growing 
radius within which pixels surrounding those flagged as cosmic rays should also be flagged. 
Cosmic ray objects in an image exhibit a non-Gaussian profile and 
therefore affect the local image statistics, causing a deviation from the 
expected Gaussian distribution of counts. The algorithm examines the histogram 
for anomalous high-count points that are separated by more than the given
threshold from the main distribution. 
The program creates a cosmic ray map and cleaned image after a specified number of iterations 
of the detection and replacement process. We note that Pych (2004) recommends 
the method as most appropriate for spectroscopic data, and less effective than 
other methods, such as those of Rhoads (2000) and van Dokkum (2001), in the case of 
images with a narrow PSF.

The final technique examined is the IRAF \textsc{xzap} task by Dickinson, a component 
of the IRAF \textsc{xdimsum} package. Unlike the other 
algorithms, this method is non-iterative, applying a spatial median filter to the 
image to perform an unsharp masking process and flagging pixels above a supplied 
threshold in a single detection step. The task will detect the background sky standard 
deviation, though the process requires manual examination of the image and 
identification and input of a section of the images that contains only sky background 
and is free of objects and cosmic rays. Other free parameters when applying the task to 
an image are the median filter box size, the cosmic ray object `zapping' 
threshold, the threshold value for object identification and the cutoff for 
sky pixel rejection. Several parameters are also available to cause pixels surrounding 
detected cosmic rays to be flagged in addition, and for setting pixels surrounding objects 
as a buffer around the object.  A map of the detected cosmic rays and a cleaned image are 
produced.

It is worth mentioning several other CRR algorithms that, though perhaps accessible, 
were omitted from the study. The IRAF task \textsc{cosmicrays} (Wells \& Bell, 
http://star-www.rl.ac.uk/iraf/ftp/iraf/docs/clean.ps.Z) of the \textsc{ccdred} 
package detects pixels above a specified threshold and flags 
them as cosmic rays if they match a flux ratio criteria relative to surrounding 
pixels. Effective application of the process is time-consuming, requiring 
either an interactive examination process to determine the appropriate flux 
ratio, or prior identification and input of the nature and location of a sample 
of image objects so that the parameter can be determined automatically. In initial 
testing, the performance of this algorithm did not match that of the other 
candidate processes closely enough to warrant more rigorous assessment. Pych 
(2004) also notes its poor performance relative to his own algorithm and an
inability to remove multiple pixel cosmic ray events.\\
Trainable classifier methods have also been omitted from the study, because the reliance on 
a training set adds an element of subjectivity in its selection, and requires 
time to obtain. 
\textsc{xnzap} (http://www.starlink.ac.uk/iraf/ftp/iraf/extern-v212/xdimsum/xdimsum.readme)
is another method included in the IRAF \textsc{xdimsum} package. 
It operates very similarly to \textsc{xzap}, but utilises an averaging filter 
rather than a median filter. It was decided to select only one of the two tasks for 
testing, and the choice was dictated largely by the greater popularity of the median 
filtering method and the results of preliminary tests.\\
Discriminating use of the {\sc SExtractor} package (Bertin \& Arnouts 1996) is another 
conceivable means of detecting cosmic rays. {\sc SExtractor} builds a catalogue 
of objects in an image, selecting candidates based on a set of detection criteria such as 
a threshold intensity, area on the CCD, and various deblending properties. Whilst promising,
developing the required technique is outside the scope of this study.     
   
\subsection{Mock data}
\label{sec:mock}
In order to equitably compare the selected CRR methods the IRAF \textsc{artdata} 
package is used to generate a controlled dataset of artificial images. Stars and 
galaxies are simulated with the \textsc{starlist} and \textsc{gallist} tasks, 
respectively, the images are compiled with \textsc{mkobjects}, and \textsc{mknoise}
is used to add noise and cosmic rays events (of various `morphologies'). 
The simulated images are 2048x2048 pixels in size, with Poisson noise added to 
an average background of 1000 counts, producing a standard deviation in the 
background of 22.5 counts. Each image has a gain of 2, a 5 electron 
read noise and FWHM of seeing of 4 pixels. The cosmic ray objects are first combined 
with galaxy and/or star objects in the datasets and other noise properties are added 
as the last step in creating an 
image.  
\textsc{mknoise} generates cosmic ray objects in an image with an 
elliptical Gaussian profile defined by the following input parameters: a random 
intensity (with a maximum of 30000 counts in our images), the radius at FWHM (in 
pixels), the minor to major axial ratio, and the position angle. The cosmic rays are 
added iteratively, varying the input parameters, in order to generate a realistic 
range of cosmic ray shapes, sizes and orientations. Appropriate combinations of the 
cosmic ray morphology parameters are generated by selecting the geometric value for
each event randomly from one of the categories in Table~\ref{tab:CRcat}. 
In each image to which cosmic rays are added, the frequency of events from 
categories 1, 2 and 3 is in the ratio 2:1:1 respectively (derived from
an inspection of cosmic ray events in our real data). 
Each cosmic ray is 
oriented with a random position angle between 1$^{\circ}$ and 180$^{\circ}$.

\begin{table}
\begin{center}
\caption{\small{The radius and axial ratio for the 3 categories of cosmic rays added 
to the artificial images.}}
\label{tab:CRcat}
\vspace{0.25cm}
\begin{tabular}{l *{3}{c}}
\hline
  Category            &  Radius range   &  Axial ratio range \\
                      & (pixels) & \\
\hline
  1 (small)            & 0.4 -- 1.0              &  0.01 -- 1.00       \\
  2 (non-elongated)   &  1.00 -- 2.00            &  0.10 -- 0.40       \\
  3 (elongated)       &  2.0 -- 5.0              &  0.005 -- 0.100     \\
\hline
\end{tabular}
\end{center}    
\end{table}

\subsubsection{Test Methods}

\begin{figure*}
\begin{minipage}{\textwidth}
\begin{center}
\subfigure[Poisson noise background: empty field]{\label{fig:mock-a}\includegraphics[width=50mm]{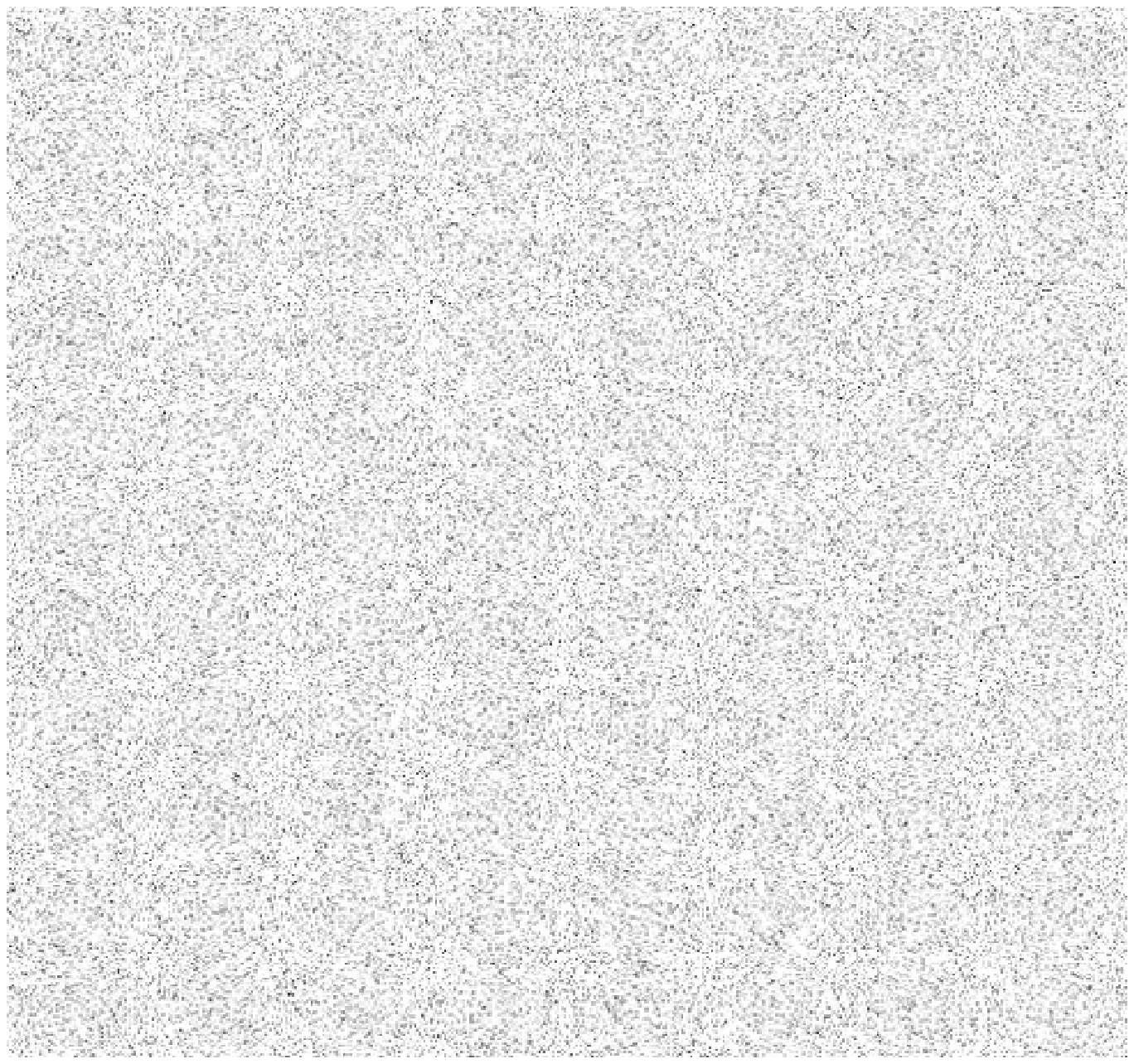}}
\subfigure[Intermediate object density: sparse galactic and stellar field]{\label{fig:mock-b}\includegraphics[width=50mm]{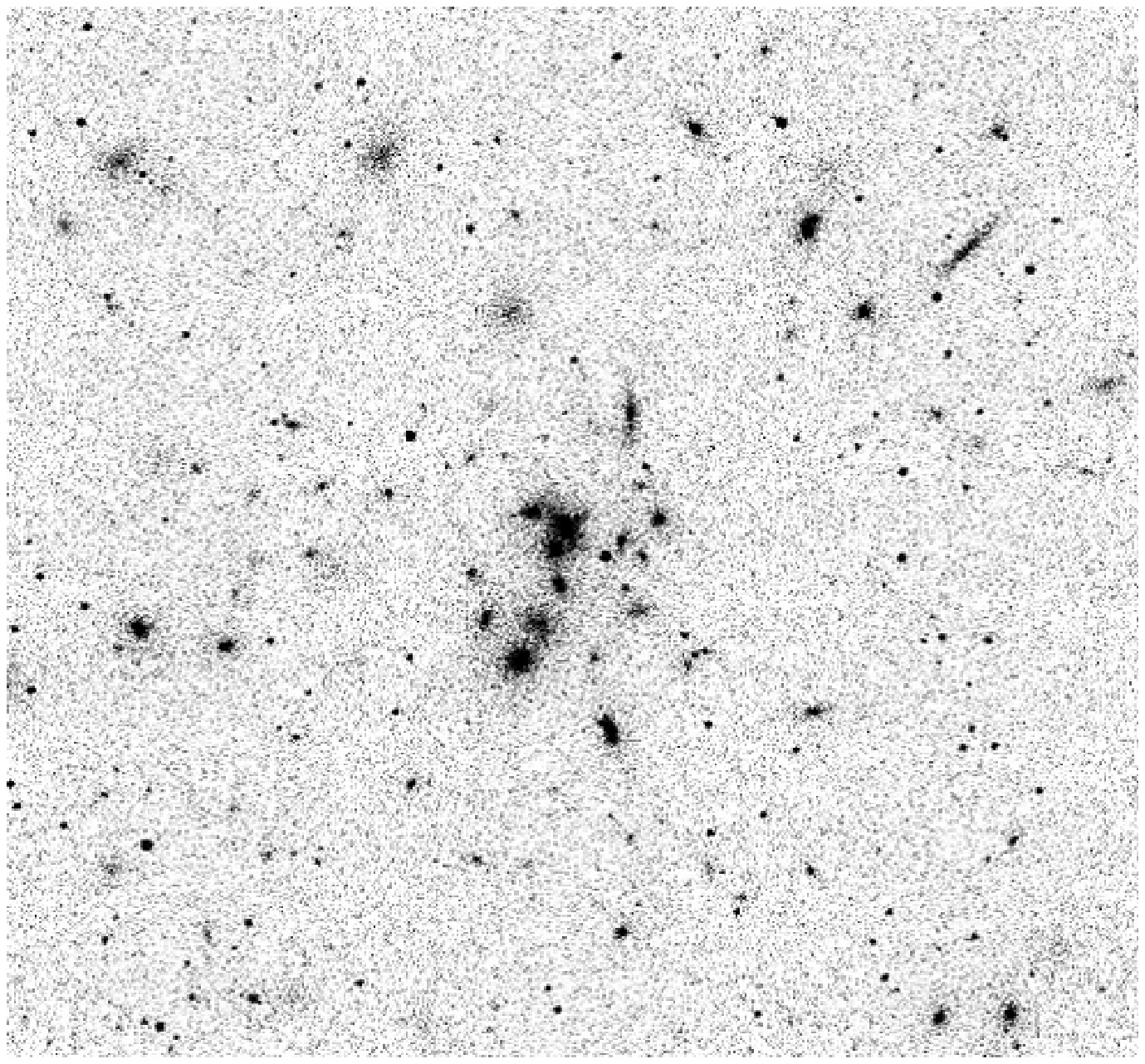}}
\subfigure[High object density: globular star cluster]{\label{fig:mock-c}\includegraphics[width=50mm]{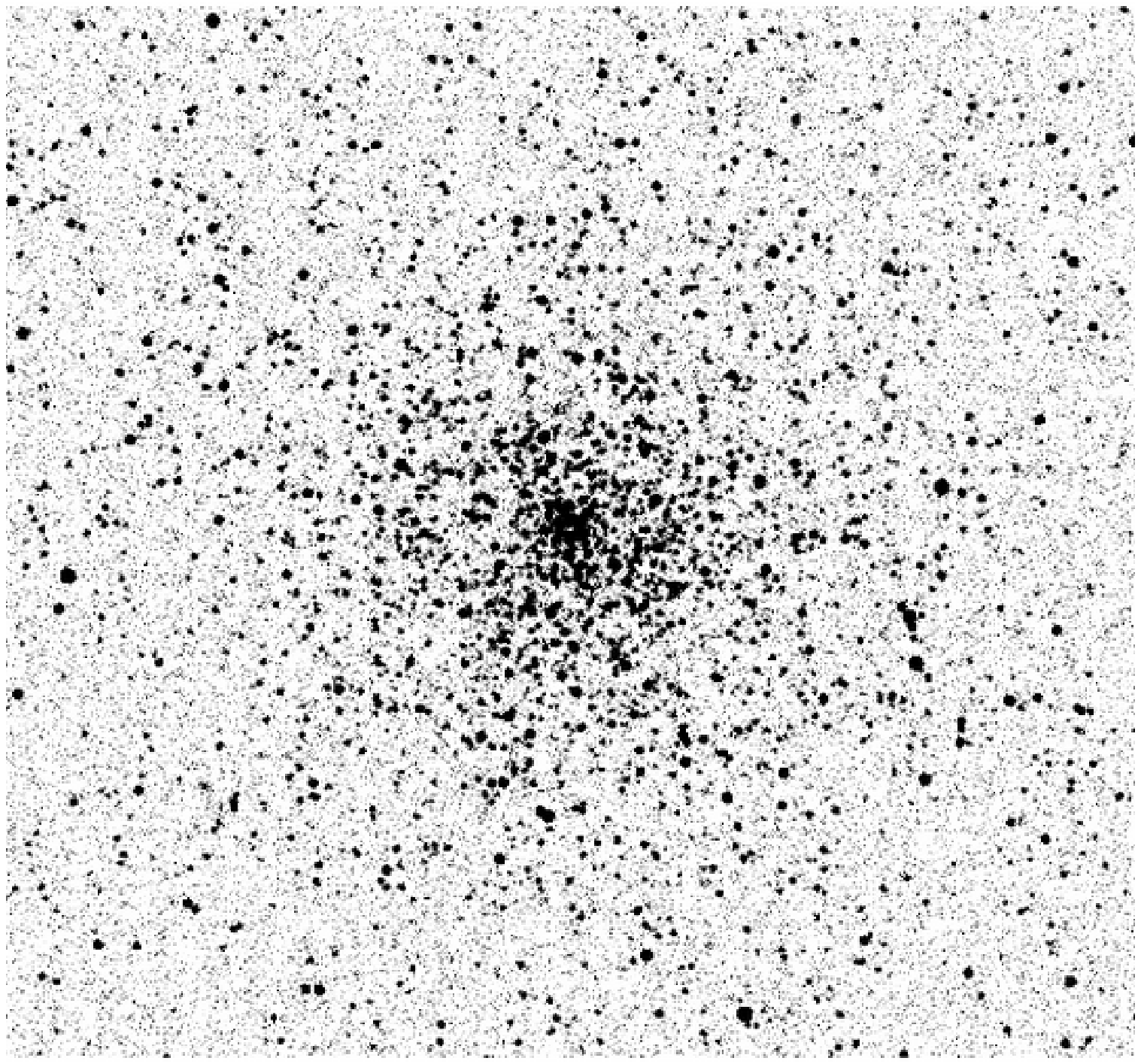}}
\end{center} 
\caption[Artificial data]{\small{Central sections of the mock images demonstrating three degrees of object density.}}
\label{fig:mock}
\end{minipage}
\end{figure*}

We identify two independent image properties that are relevant to the investigation 
of CRR techniques and with which we can describe a particular astronomical image: 
\begin{enumerate}
\item the density of real (non-noise) objects in the field; and 
\item the frequency of cosmic ray affected pixels. 
\end{enumerate}
Evidently, the CRR task should become more difficult 
as either of these properties increase 
and provide greater opportunity for intersection between cosmic rays and objects and 
loss of data to impinging cosmic rays. A qualitative measure of the former attribute 
is used in developing the mock dataset. Three images are created that exemplify three 
distinct and disparate degrees of object density as shown in Figure~\ref{fig:mock}. 
The first, Figure~\ref{fig:mock-a}, is the trivial case of an image consisting simply 
of a uniform noise background, the second, Figure~\ref{fig:mock-b}, a field containing 
a number of sparse stellar and galactic objects and the third, Figure~\ref{fig:mock-c}, 
an extremely crowded image containing a large globular star cluster.

The frequency of cosmic ray events is quantified using a cosmic ray `filling 
factor'. This is defined as the percentage of image pixels that are part of a 
cosmic ray event. For the simulated cosmic rays, such pixels are defined to be 
those that, after embedding the cosmic ray image in a noise background, have an 
intensity $\ge 4\sigma$ above the mean background level. Eleven levels 
(see Table~\ref{tab:CRff}) of cosmic ray filling factor are used in combination with the 
three base images to create an overall set of 33 images. For each level, a 
known number of cosmic ray objects are added to the images to create an 
approximately even distribution of filling factors. The {\sc SExtractor} package 
(Bertin \& Arnouts 1996) is used to catalogue the number of cosmic ray objects that
are then present in each image, since the addition of Poisson noise obscures the 
weaker and smaller cosmic rays added. A mask image is created for each level 
indicating the location of all cosmic ray affected pixels (with intensity $\ge 
4\sigma$ and excluding `hot' background pixels). The number of cosmic ray pixels can 
then be counted and expressed as a percentage of the total image pixels to give the 
cosmic ray filling factor for each level, as shown in Table~\ref{tab:CRff}. The 
dependance of the final value on the random distribution and morphologies of the 
events prevents us arbitrarily selecting evenly spread, round numbers for the filling 
factor variable.   

\begin{table}
\begin{center}
\caption{\small{Cosmic ray filling factor levels used in mock data trials.}}
\label{tab:CRff}
\vspace{0.25cm}
\begin{tabular}{ c c }
\hline
  Filling factor  &  N(cosmic rays) \\
  (per cent) & \\
\hline
  0.06  &  233  \\
  0.12  &  463  \\
  0.24  &  918  \\
  0.37  &  1357  \\
  0.49  &  1801  \\
  0.61  &  2232  \\
  0.73  &  2675  \\
  0.86  &  3117  \\
  0.98  &  3550  \\
  1.10  &  3986  \\
  1.22  &  4412  \\
\hline
\end{tabular}
\end{center}    
\end{table}

Selecting the optimal settings of the free parameters for a given method of 
CRR in an image is a compromise between maximising the cosmic ray detections, 
minimising the number of false detections and ensuring a satisfactory cleaning 
of the image. Ideally, it would be preferable to experiment 
iteratively with the settings of an algorithm for each cosmic ray-affected image, 
to determine the combination of parameters that produces results of the quality 
required for an intended application. This is a prohibitively time consuming 
process, in the case of a time series analysis, where a large number of images need 
to be processed. 
In this investigation however, given prior knowledge of the details of the cosmic 
rays in the image and to provide a fair comparison of the methods tested, 
experimentation is extensive. The CRR procedures are repeatedly applied to a 
selected artificial image of intermediate object density (Figure~\ref{fig:mock-b}) 
and $\sim$0.7 per cent filling factor (Table~\ref{tab:CRff}), as the relevant task 
inputs are iteratively varied. 
The detection pixel map produced in each instance is examined to determine the number 
of flagged pixels known to be true cosmic rays and, as a result, the number of falsely 
flagged pixels. 
The process is defined to be `optimised' (at least for this image) by the parameter 
set resulting in the maximum number of genuine cosmic ray pixel detections with a 
false detection rate of less than 0.01 per cent. 
The appropriate parameter set for each algorithm is then used to process each image 
of the mock dataset and the outcomes are used as the basis for comparison of 
the performance of the processes. In this way we impose a consistent standard 
on each algorithm and evaluate at a similar level of performance across the dataset, 
under the assumption that similar noise properties among the images will 
ensure reasonable results in each case. 

For those algorithms that produce a cleaned version of the image, another 
important aspect of the performance is the quality of this image reconstruction. 
Explicitly, we are concerned with how well the value of flagged cosmic ray pixels 
are interpolated from the real data, the degree to which cleaning of false detections 
affects the image data and whether artifacts of the cleaning process can be 
observed in the output image. 

\subsection{Real data}
The capabilities of the algorithms are also evaluated using a real image that contains  
true cosmic rays incurred during the exposure. The observation was made with the Wide 
Field Imager (WFI) on the Anglo-Australia Telescope, which consists of an 8k~x~8k 
array of 8 CCD panels. The mosaic image that is the subject of tests contains sparse
stellar and galactic objects and a significant sample of true cosmic rays. The FWHM of 
seeing is 5.3 pixels.
In addition to the genuine noise events present, a sample of artificial cosmic rays 
is added to the images and the CRR performance is assessed by a blind test. The 
success of detection and removal of the simulated cosmic rays is taken to be indicative 
of the efficiency of real cosmic ray removal. 600 events with morphologies similar to 
that described in Section~\ref{sec:mock} above are added randomly across the image, 
and a cosmic ray map is created of all pixels affected by the artificial cosmic rays so 
that detections of artificial objects can be examined. Cosmic ray pixels are again defined 
as those containing flux $\ge 4\sigma$ above the mean background level.
These tests are more qualitatively conducted than the artificial image testing presented in 
the previous section. The algorithms are applied several times to each image, making 
manual adjustments to the free parameters in an attempt to customise the process to the 
image. The best result is selected by visual inspection of the output.

\section{Results}
\label{sec:results}

\subsection{Mock Data}

Using various manipulations of the output images (cosmic ray maps and cleaned images) 
with the artificial cosmic ray mask images, it is possible to quantify the following 
properties of the results for each subject image: 
\begin{enumerate}
\item The number of true cosmic ray pixels that are \mbox{detected; and}
\item The number of pixels that are falsely flagged as cosmic rays (e.g. pixels containing 
only object or improbably high background flux).
\end{enumerate}

\begin{figure*}
\begin{minipage}{\textwidth}
\begin{center}
\subfigure[Detection efficiencies of the Rhoads (2000) and \textsc{xzap} algorithms.]{\label{fig:det-RhXz}\includegraphics[width=8.5cm]{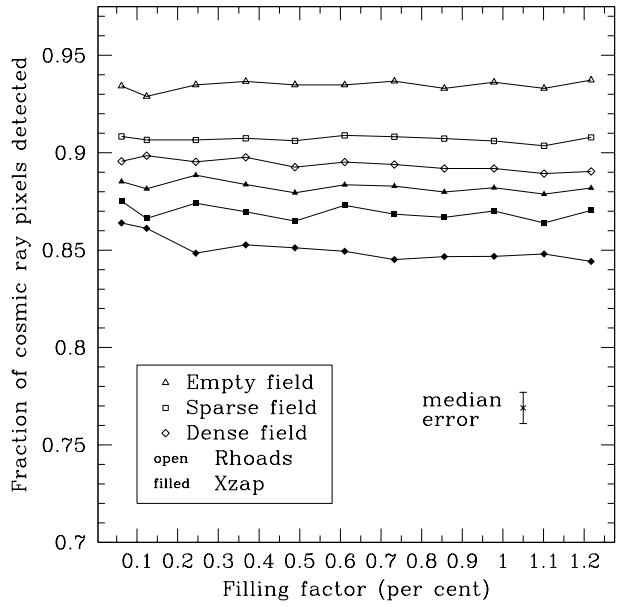}}
\subfigure[Detection efficiencies of the Pych (2004) and van Dokkum (2001) algorithms.]{\label{fig:det-PyDk}\includegraphics[width=8.5cm]{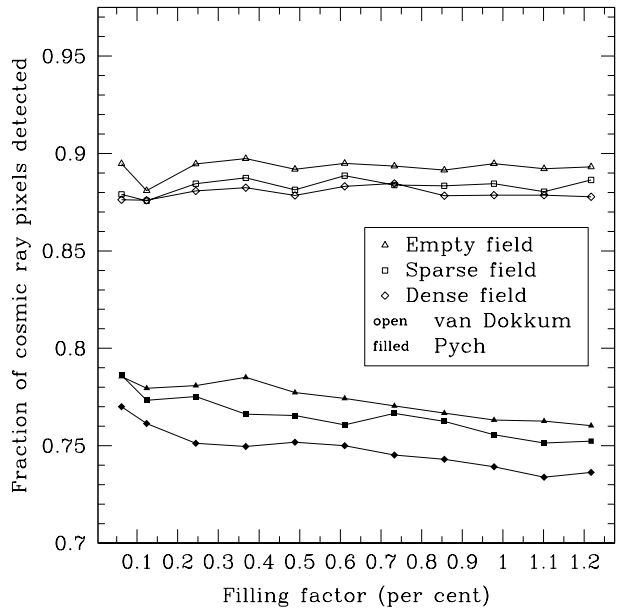}}
\end{center} 
\caption[]{\small{Cosmic ray pixel detection efficiency results. For images of similar object density, the effiency of detection is independent of the cosmic ray filling factor, though reduced by increasing object density. Poisson errors are indicated as a median error magnitude.}}
\label{fig:det}
\end{minipage}
\end{figure*}

The value of (i) is taken to be the best indicator of the efficiency of detection of 
cosmic ray events. The results are expressed as the fraction of the 
total number of cosmic ray pixels in the original image (those contributing to the image 
filling factor). 
Figure~\ref{fig:det} presents the detection efficiency results from the mock data trials 
decribed in Section~\ref{sec:mock}, for the four algorithms as a function of the cosmic 
ray filling factor. 

\par
\begin{figure*}
\begin{minipage}{\textwidth}
\begin{center}
\subfigure[van Dokkum algorithm]{\label{fig:fp-Dk}\includegraphics[width=7.5cm]{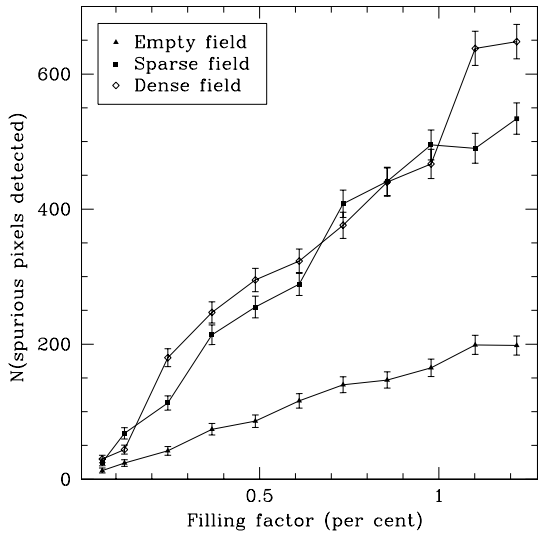}}
\subfigure[Pych algorithm]{\label{fig:fp-Py}\includegraphics[width=7.5cm]{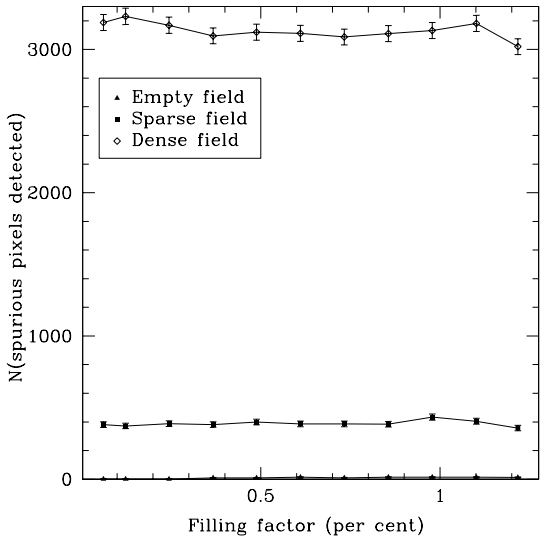}}
\subfigure[Rhoads algorithm]{\label{fig:fp-Rh}\includegraphics[width=7.5cm]{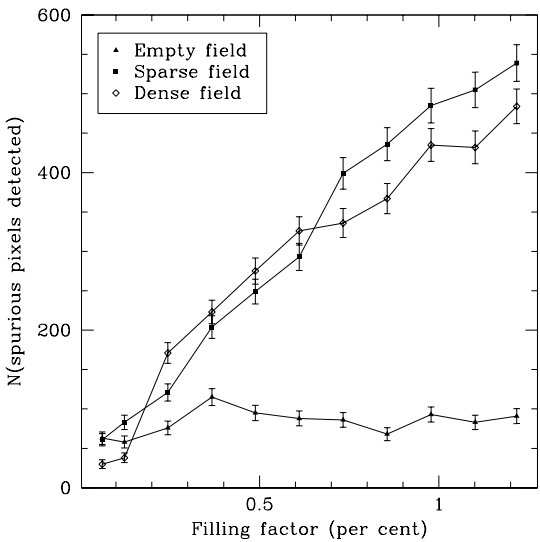}}
\subfigure[{\sc xzap} algorithm]{\label{fig:fp-Xz}\includegraphics[width=7.5cm]{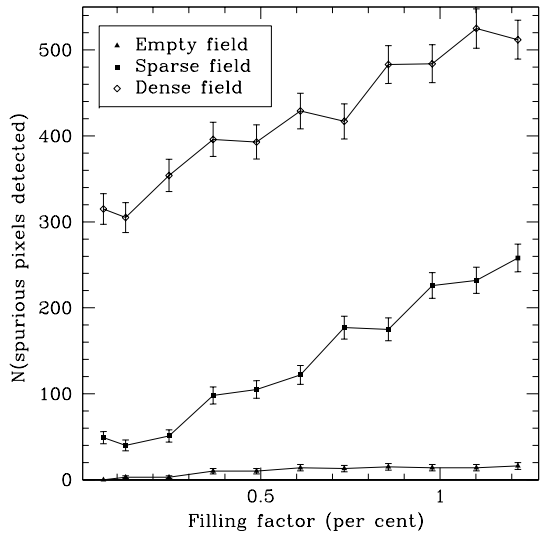}}
\end{center} 
\caption[]{\small{The number of spurious detections made by each algorithm as a function of 
filling factor, with simple Poissonian errors. Unlike the detection efficiency, the rate of 
false detections appear to have some dependency on the cosmic ray density for all other than 
Pych's algorithm. The adverse effect of increasing object density is again demonstrated in 
these results.}}
\label{fig:fp}
\end{minipage}
\end{figure*}

The false detections made during the CRR process, (ii), represent real data that is altered 
by the algorithm. We wish to minimise this data loss, of which we would 
generally be unaware of unless the results are thoroughly examined by eye.
Figure~\ref{fig:fp}  shows the numbers of spurious detections made by the algorithms 
during the testing of simulated images. These results, as a function of the filling factor, 
are presented individually for the different algorithms such that the effect of image properties 
can be more clearly observed. It is seen that comparable levels of false detections are made 
amongst the test subjects, with the exception of the Pych algorithm in the high object density limit.  

\subsection{Verification}

The efficiency and false detection results can be confirmed using the reconstructed clean 
image if one is output by the algorithm. Of those tested, the van Dokkum, Pych and 
\textsc{xzap} algorithms produce a cleaned image as direct output so that assessment of the quality 
of cleaning is applicable.   
Analysis of the cleaned image confirms that the van Dokkum and Pych processes clean only those 
pixels that are flagged; that is, the number original image pixels that are altered is equal to 
the number of pixels indicated as detections in the output cosmic ray map. The Pych algorithm 
provides an option to flag pixels surrounding detections (and hence clean them), however these 
options are not utilised in order to avoid the resulting spurious detections that would be 
associated with the cosmic ray events.
Notably, the \textsc{xzap} task is found to clean more pixels than are flagged, 
although again, the parameters that specify neighbouring pixels to be flagged or altered 
have not been applied. Thus, more of the actual cosmic ray pixels are altered and the 
`cleaning efficiency' is approximately 2 percentage points higher than is indicated by the 
detection efficiencies of Figure~\ref{fig:fp-Xz}. Examining the reconstructed images also then 
reveals a greater number of false detections surrounding detected cosmic ray pixels.

\subsection{Real Data}

The results of the blind tests on the real data appear in Table~\ref{tab:real}. These are the 
best results that could be obtained with limited experimentation. The large image size brought 
to notice the issue of time constraints, and here the Pych and \textsc{xzap} processes gain 
some advantage in requiring far less operating time than either the Rhoads or van Dokkum 
algorithms, which are 
relatively computationally expensive. The van Dokkum algorithm, however, 
clearly produces the best detection performance.  
This is in contrast to the mock data results.  
We consider that there may be an over abundance of larger
cosmic ray events in our mock data. 
We suggest that this contributes to decreased performance
of the van Dokkum algorithm on the mock data, 
since it is observed that events missed by this
method are generally the largest cosmic ray events present.

\begin{table}
\begin{center}
\caption{The best fractional cosmic ray efficiencies achieved for the real data set. 
Errors are simple Poissonian errors.}\label{tab:real}
\begin{tabular}{l *{4}{c}}
\hline 
CRR Method & N(detections) & Error      & Efficiency  & Error       \\
           & (pixels)      & (pixels)   &             &             \\
\hline 
van Dokkum &   5969        & 77         &  0.86       &  0.015       \\
Rhoads     &   5410        & 74         &  0.78       &  0.014       \\
Xzap       &   5405        & 74         &  0.78       &  0.014      \\
Pych       &   5267        & 73         &  0.76       &  0.014       \\
\hline
\end{tabular}
\medskip\\
\end{center}
\end{table}

\section{Discussion and Conclusions}
\label{sec:conc}
It is interesting to note that the filling factor appears to have little or no effect on the 
performance of cosmic ray detection (Figure~\ref{fig:det}). Only the Pych results show even 
a noticeable drop in the efficiency of detection with increasing filling factor,in the 
range of values tested. 
This is contrary to our intuitive prediction that the detection efficiency would be 
significantly affected by the larger numbers of intersecting events 
% that lack a standard cosmic ray morphology for detection 
as the filling factor is increased.
Also from Figure~\ref{fig:det} the image object density is of greater importance, causing 
significant reductions in performance as it is increased. This is most noticable in the 
results from the Rhoads algorithm, which are markedly better for the empty, 
noise image. As would be expected, it is also clearly illustrated by the effects on the 
rate of spurious detections; in particular by the Pych algorithm results 
(Figure~\ref{fig:fp}), which exhibit a 
dramatic increase in false detections in the most dense image field where there is a much 
larger number of candidate objects for false dectection.  

In terms of detection, the algorithms do not perform as well as expected in the tests of the 
empty, noise image. It is presumed that a much better result would be obtained for the trivial 
case by determining the optimising parameter set using the empty image, eliminating the 
need for the algorithm to distinguish between objects and cosmic rays.

The false detection rates, with the exception of the results from Pych's algorithm, appear to 
exhibit some dependence on the filling factor, in addition to being affected 
by the object density. Unlike the effects of increasing the object density which, in providing more 
objects that could potentially be falsely flagged as cosmic rays, might be expected, the increase 
in false detections with filling factor is unexpected and counter-intuitive. It is explained, 
however, by comparison of false detection image maps from each algorithm with the original cosmic 
ray maps created for addition to the simulated images, prior to the addition of background noise 
and imposing the definition of cosmic rays as objects with intensities $\ge 4\sigma$. It is clear 
the vast majority of false detections made by the van Dokkum and Rhoads algorithm and most of those 
made by the \textsc{xzap} algorithm are sub-threshold simulated cosmic rays or pixels at the edges 
of simulated cosmic rays. These pixels are then not true false detections, but instances where  
the addition of noise to the image with added cosmic rays has 'smeared' the edges and low-intesity 
cosmic rays to below the subsequently defined cosmic ray threshold. The Pych algorithm results 
do not exhibit the same detection of sub-threshold cosmic rays and so the results in 
Figure~\ref{fig:fp-Py} show no filling factor dependency. In hindsight, a more successful 
method would remove subthreshold pixels from cosmic ray before adding them to the 
artificial images. An indication of the numbers of false detections made by the van Dokkum, 
Rhoads and \textsc{xzap} algorithms can be better derived from the values detected at very low 
filling factors.
 
\begin{figure*}
\begin{minipage}{\textwidth}
\begin{center}
\subfigure[Original image]{\label{fig:cr7}\includegraphics[width=3.7cm]{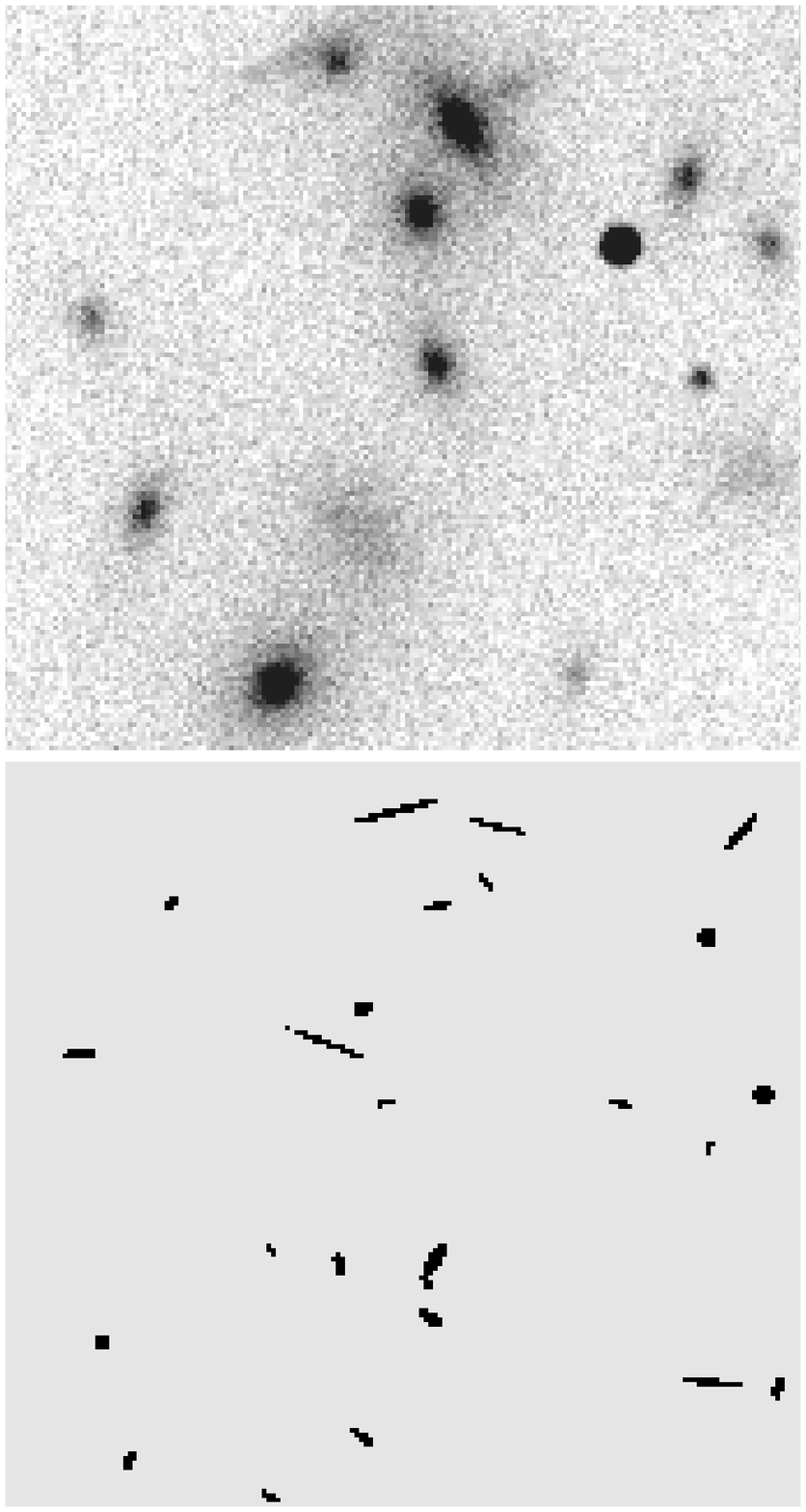}}
\subfigure[van Dokkum algorithm]{\label{fig:cr7dk}\includegraphics[width=3.7cm]{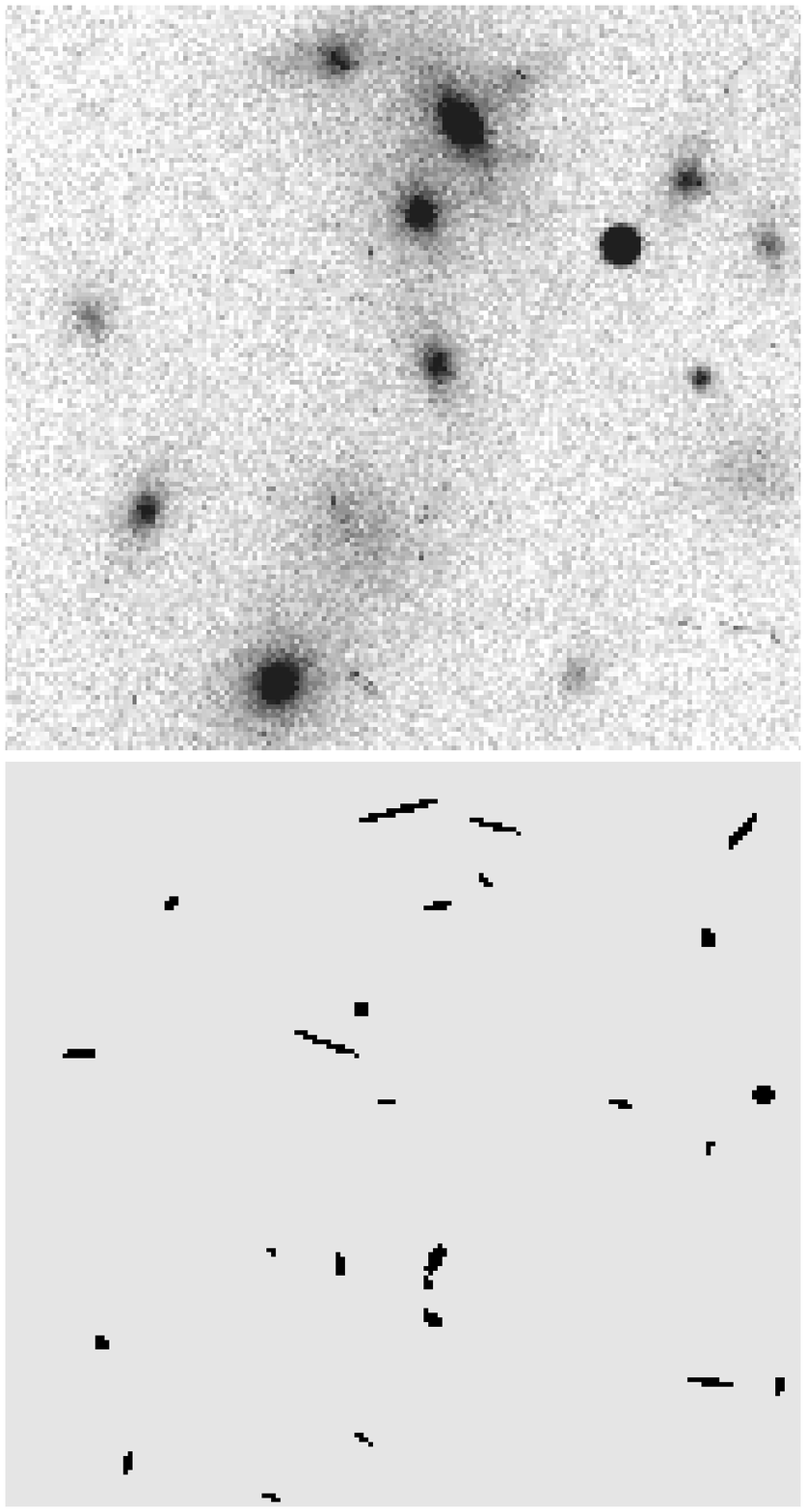}}
\subfigure[Pych algorithm]{\label{fig:cr7py}\includegraphics[width=3.7cm]{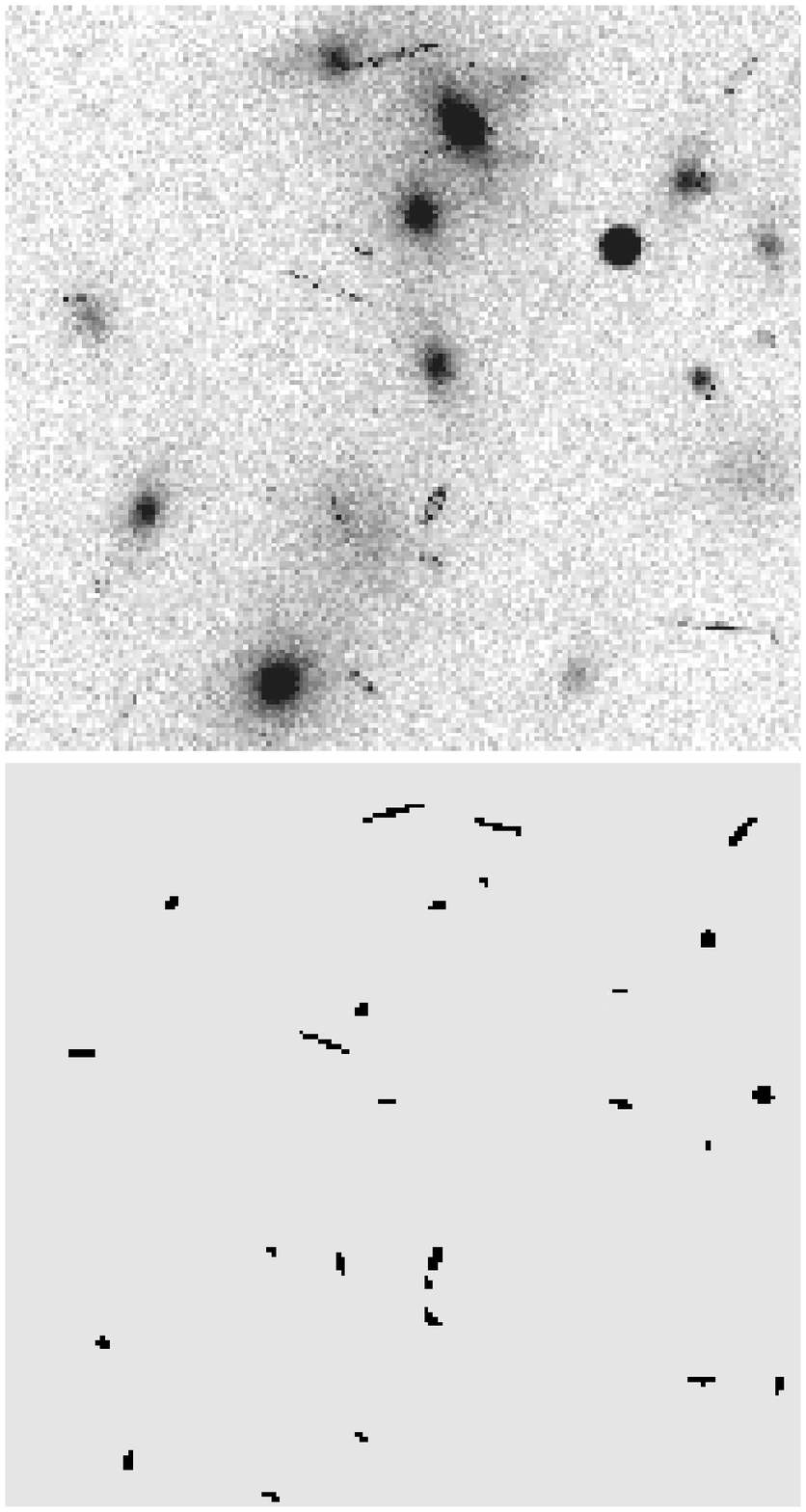}}
\subfigure[{\sc xzap} algorithm]{\label{fig:cr7xz}\includegraphics[width=3.7cm]{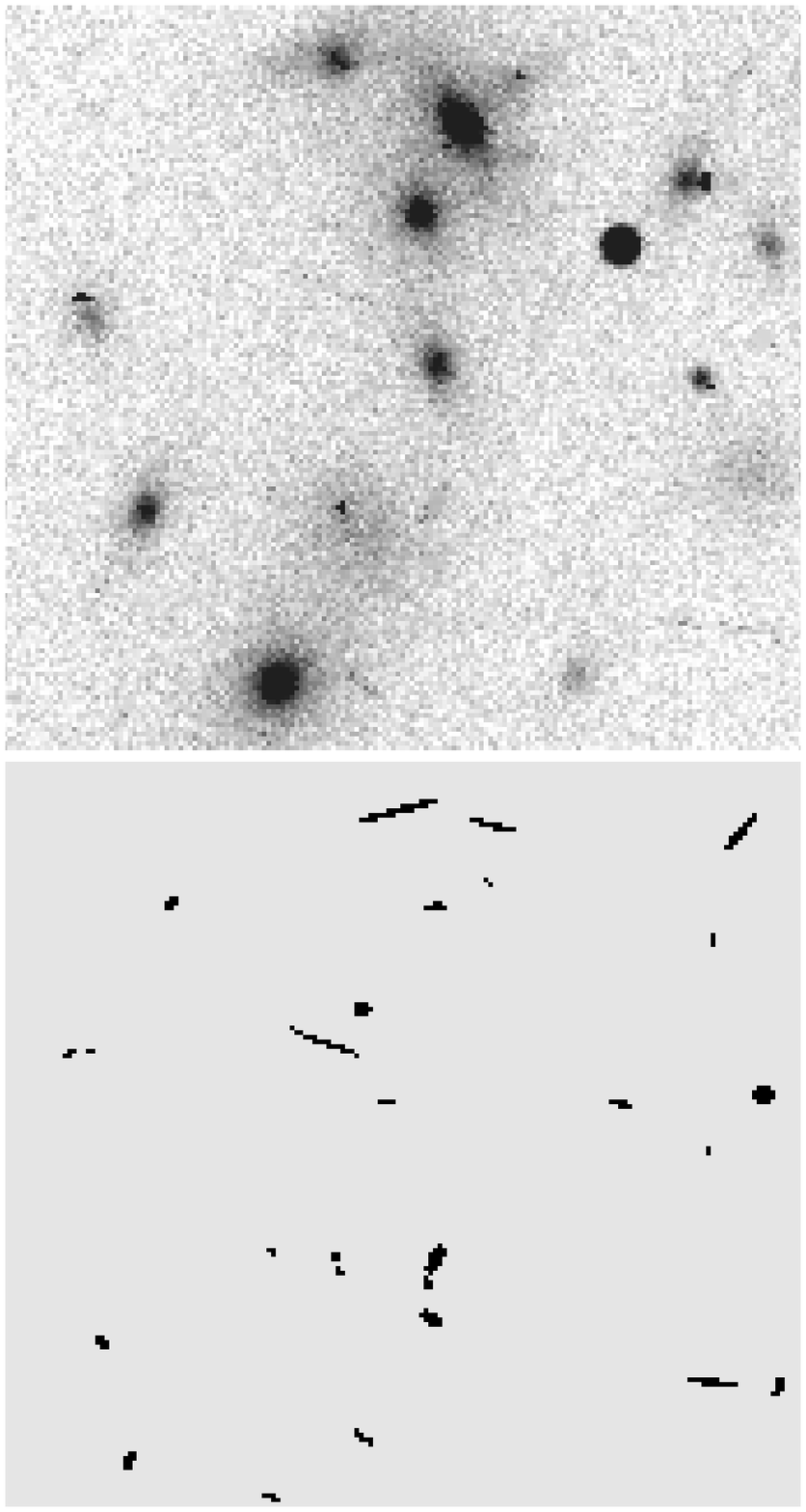}}
\end{center}
\caption[]{\small{A sample of resulting cleaned and cosmic ray map image sections for the 
artifical image that is used to optimise the algorithm performance in the tests.}}
\label{fig:cr7eg} 
\end{minipage}
\end{figure*}

The quality of the image cleaning is best illustrated by example, since it is somewhat objective 
and is largely judged by the visual appearance. 
Figure~\ref{fig:cr7eg} compares a portion of the test image, of intermediate object and 
cosmic ray density, and the reconstructed image output by the three algorithms that 
produce a cleaned image. 
The original image, prior to cosmic ray addition, is at top in Figure~\ref{fig:cr7}, with 
the true cosmic ray image beneath. Figures~\ref{fig:cr7dk}, \ref{fig:cr7py} 
and \ref{fig:cr7xz} show the cleaned output image above and the detected cosmic ray map 
below from the van Dokkum, Pych and {\sc xzap} algorithms, respectively.

From general observations of the results obtained throughout the study, the van Dokkum 
algorithm produces a very well-cleaned image, though the algorithm tends 
sometimes to miss detecting the relatively larger and less elongated cosmic ray events.
The cleaned output of the \textsc{xzap} algorithm is also impressive in many cases, 
though the use of a small median filter box (to reduce detections of true objects) can 
cause only sections of cosmic rays to be detected, and hence some remnant to remain in 
the image. Though the Rhoads algorithm does not directly output a cleaned image, the 
algorithm sets the highest standard of detection in the mock data trials, and performs 
well on the real data tested.
Pych's (2004) advice that his algorithm is designed for spectroscopic, rather 
than imaging, data should perhaps be heeded, as the alorithm's performance on the 
imaging data tested was certainly not equal to that of the others. However, it must be 
emphasised that since no comparative testing has been performed on spectroscopic data, and
the Pych algorithm performance may well exceed that of other methods here. The analysis of the 
image in small subframes often produces spurious detections of objects at the frame edges 
and sometimes leaves artifacts in the image, such as anomolous alteration of data at the 
edges or outer regions of multiple-pixel cosmic rays.   
All the algorithms make spurious detections at regions of overexposure on the CCD, and at 
regions of non-uniformity in the background flat-fielding, such as striations where the mean 
background levels are significantly altered to higher intensities.

In conclusion, the primary findings of the investigation are summarised below:
\begin{enumerate}
\item The cosmic ray detection efficiencies and false detection rates of the CRR algorithms 
tested are independent of the density of cosmic ray events in an image (in the range of 
`filling factors' tested), but are affected by the density of true objects native to the field 
observed. 
\item From the mock data trials, Rhoads' detection algorithm exhibits the highest cosmic ray 
detection efficiency of those tested, and exhibits a reasonable (less than 0.02 per cent) false 
detection rate in mock data trials. It is followed closely by the van Dokkum algorithm, which 
actually supercedes the results of Rhoads' algorithm when applied to the real observational data 
sample described. 
\item The van Dokkum algorithm produces the most satisfactory cleaned image as output, though 
it is a relatively slow process. The IRAF \textsc{xzap} algorithm provides a faster alternative, 
at a cost of a decrease in the quality of the result.
\end{enumerate}

Further investigation and testing of both a more extensive set of images types (particularly 
to include spectroscopic data) and cosmic ray rejection methods would be prudent to elaborate 
on and extend the results presented here.

\section*{Acknowledgments}
CLF was supported by a vacation scholarship throughout the course
of this work. 
KAP acknowledges support from an EPSA University of Queensland 
Research Fellowship and a UQRSF grant.

\end{document}